\documentclass[pre,a4paper,floatfix,twocolumn]{revtex4}
\usepackage{graphicx}
\usepackage{amsmath}
\usepackage{mltex}
\ifx\pdfoutput\undefined
\else
  \ifx\pdfoutput\relax
  \else
    \ifcase\pdfoutput
      \DeclareGraphicsExtensions{.eps}
    \else
      \DeclareGraphicsExtensions{.pdf}
    \fi
  \fi
\fi

\renewcommand{\deg}{{}^{\circ}}
\renewcommand{\phi}{\varphi}

\begin{document}

\title{Erosion patterns in a sediment layer}
\author{Adrian Daerr}
\email{daerr@ccr.jussieu.fr}
\altaffiliation[permanent address: ]{MSC/Univ. Paris VII/PMMH-ESPCI, 10 Rue
  Vauquelin, F-75231 Paris cedex 05}
\author{Peter Lee}
\email{lee@phys.leidenuniv.nl}
\altaffiliation{Department of Physics, Leiden University, Niels 
Bohrweg 2, NL-2333 RA Leiden} \author{Jos\'{e} Lanuza}
\author{\'{E}ric Cl\'{e}ment}
\email{erc@ccr.jussieu.fr}
\affiliation{Laboratoire des Milieux D\'esordonn\'es et \\
H\'{e}t\'{e}rog\`{e}nes, UMR7603 - Universit\'{e} Pierre et Marie Curie - Bo\^{\i }te 86,\\
4 place Jussieu, 75252 Paris CEDEX 05, France}
\date{\today}

\begin{abstract}
  We report here on a laboratory-scale experiment which reproduces a
  rich variety of natural patterns with few control parameters. In
  particular, we focus on intriguing rhomboid structures often found
  on sandy shores and flats. We show that the standard views based on
  water surface waves do not explain the phenomenon and we
  evidence a new mechanism based on a mud avalanche instability.
\end{abstract}

%
%

\maketitle

Many patterns observed \cite{allen1984,dodds2000} in
natural environments stem from erosion/deposition processes. These
structures are related to transport of solid granular particles via a
fluid phase that can be either a gas, a liquid, or even a flowing
granular phase. They span a huge variety of spatial and temporal
scales. Examples of these are fractal river basins \cite{dodds2000},
meandering rivers \cite{edwards2002}, dune fields \cite{kroy2002},
granular avalanches \cite{daerr2001pof}, ripple marks \cite{betat1999}
on sand banks or on coastal continental platforms. Due in part to its
economical and environmental impact, elementary transport processes
involved in erosion are still the focus of intense scientific
scrutiny.

It is notoriously difficult to provide a fully consistent description
of particle laden flows either from a one phase or a two phase point
of view \cite{cargese1997}. Most of the practical knowledge on erosion
comes from empirical laws often derived from field measurements
\cite{graf1993}. Several tentatives were made recently to tackle from
the statistical physics point of view the dynamics of formation of
river basins \cite{dodds2000} but many question are raised when one
attempts to relate basic transport properties to large scale pattern
forming instabilities.

In this letter we report on an experimental setup which is designed to
produce a generic situation of a falling water level on an erodible
sediment layer. This occurs naturally when the sea retreats from the
shore or when a reservoir is drained \cite{allen1984}. We use a
plexiglass container with a flat bottom. A square
$130\,$mm$\times 130\,$mm plate of de-polished plexiglass slides along
the bottom, and is pulled by a motor through a translation stage. The
whole setup can be tilted to an arbitrary angle $\theta $. The first
step of the experiment is the deposition under water of a sedimented
powder layer covering the mobile plexiglass sheet. To this end, the
inclination of the whole setup is lowered so that the bottom of the
container and the sheet are horizontal ($\theta =0$). A suspension of
alumine oxide powder is then quickly poured into the container. We use
commercial abrasive powder made of rough grains with mean diameter
$d\simeq 30\,\mu $m and density $\rho\simeq 2.75\,\mbox{g/cm}^{3}$.
The liquid is demineralized water, to which we add hydrochloric acid
so as to lower the $pH$ of the suspension to about $4$ in order to
prevent flocculation of the grains.
\begin{figure}[hb!]%
  \includegraphics[width=0.95\columnwidth,height=0.8\textheight,keepaspectratio]{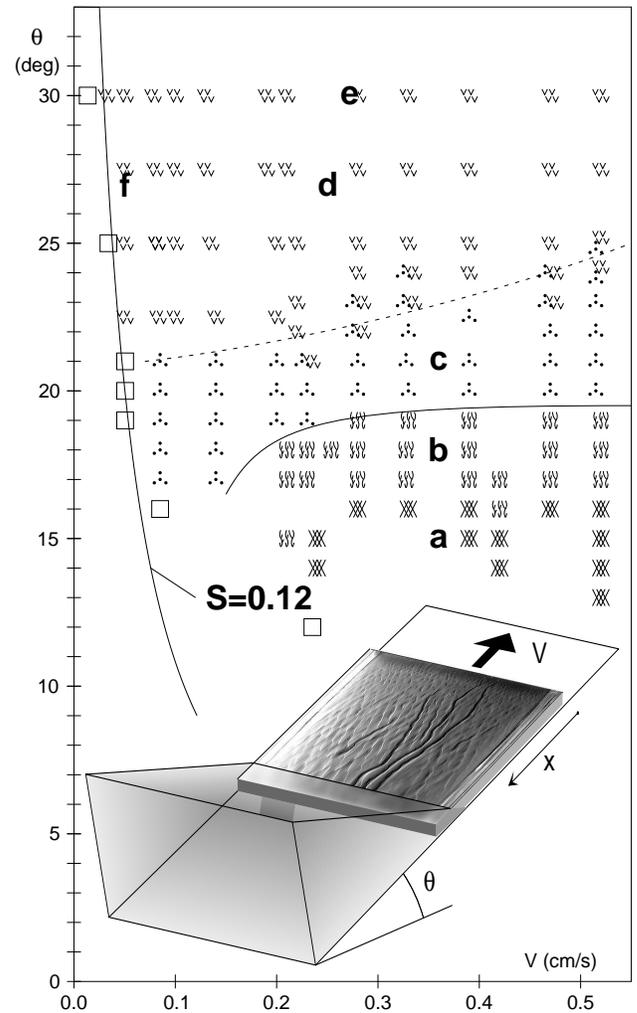}%
  \caption{Phase diagram localizing the different patterns observed in
    the erosion experiments in the tilt angle $\theta$ - velocity $V$
    space. Lines delimiting the domain boundaries are mere guides to
    the eyes. Letters and symbols correspond to
    Fig.~\ref{fig:samples}.}
  \label{fig:phase_diagram}%
\end{figure}%
\begin{figure*}%
  \includegraphics[width=\textwidth]{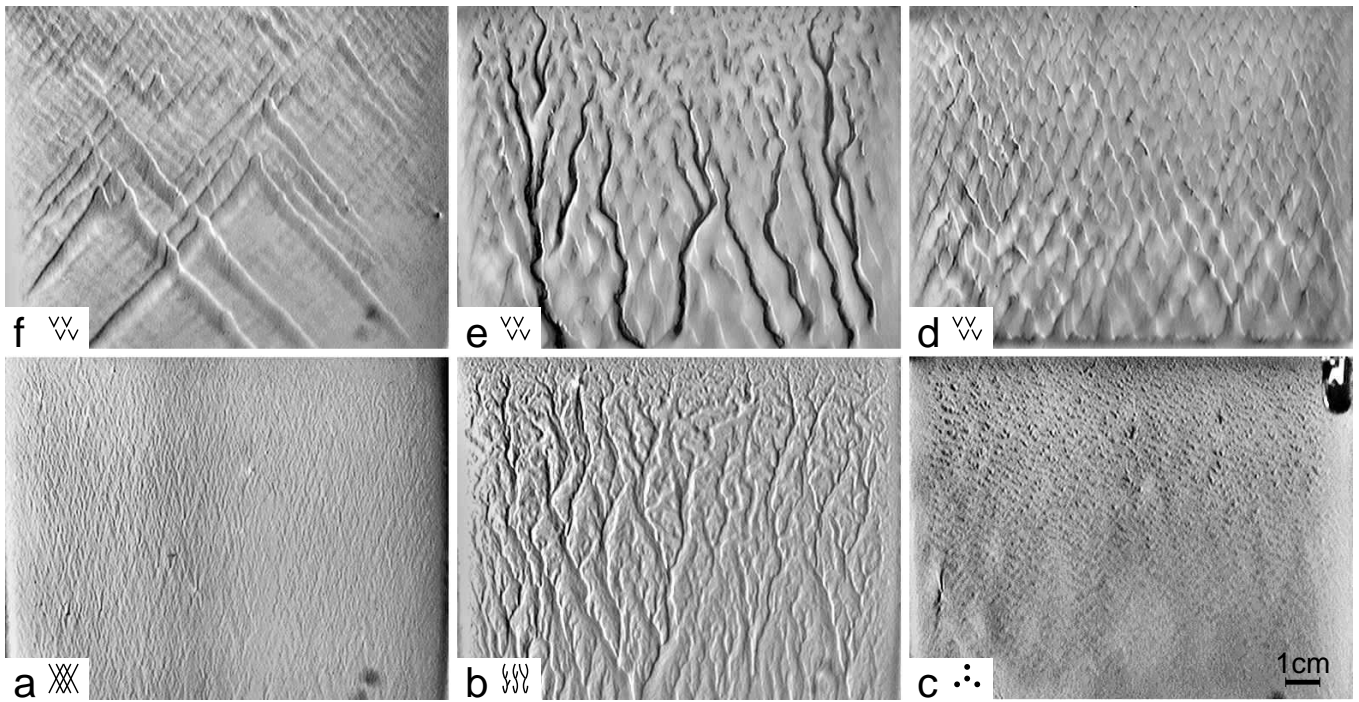}%
  \caption{Patterns observed in the erosion experiment: {\bf a} crossed
    hatched pattern, {\bf b} disordered branched pattern, {\bf c} orange skin,
    {\bf d} chevron structure, {\bf e} chevrons with oblique channels, {\bf f}
    localized pulses at chevron onset. The layer appears darker where it has 
    been eroded because the bottom plate is black. A light source to the
    left creates additional shading.}
\label{fig:samples}
\end{figure*}%

After deposition of a sedimentary layer of typical height $1\,$mm, the
container and the plate are tilted to an angle $\theta $. Part of the
liquid is then slowly drained, until the contact line of the liquid
free surface with the tilted bottom of the container reaches about
$1$cm below the top of the mobile sheet. Then the sliding plate is
pulled out of the liquid at a speed $V$ (see inset of
Fig.~\ref{fig:phase_diagram}). The layer is filmed by a CCD camera.
Its position is fixed with respect to the plate and its optical axis
is perpendicular to the surface. This simple set-up allows us to
observe a variety of phenomena and structures depending on two control
parameters, the angle $\theta $ and the speed $V$. The ``phase
diagram'' is sketched on Fig.~\ref{fig:phase_diagram}. When the
sedimented granular layer is pulled out very slowly ($V\leq
0.04$\thinspace cm/s), no pattern is observed. The liquid seeps out
of the sediment, which dries progressively without being altered.
However, above a critical speed and tilt corresponding on
Fig.~\ref{fig:phase_diagram} to empty squares, we observe the
formation of erosive patterns on the granular sediment surface. For
decreasing angle, the patterns faint away progressively and for a tilt
angle $\theta <13{{}^\circ}$, it is often difficult to witness of their
presence. 

At velocities greater than approx.\ $0.1$\thinspace cm/s, surface
structures appear clearly when the layer is tilted to more than about
$14\deg$. For angles close to this value, the draining liquid leaves a {\em
  cross-hatched}, dense pattern of very small and shallow channels
(see Fig.~\ref{fig:samples}a).

Around $18\deg $, we obtain a branched, disordered river-network,
whose biggest branches have widths of about $1\,\mbox{mm}\approx 30d$
(Fig.~\ref{fig:samples}b). After the passage of the water contact
line, the surface of the sediment is still smooth and the pattern
appears with a delay of few tens of seconds. The dynamical evolution
of the structure can last as long as two minutes. First, small
localized structures with a characteristic angle similar to the
previous cross-hatched pattern appear almost everywhere and then,
bigger and bigger disordered structures are created as they merge
under the action of erosion and sediment transport. We will call this
regime the {\em disordered regime}, refering to the random aspect of
the final network. Transition between cross-hatched and branched is
progressive.

For velocities higher than $0.1\,$cm/s and slopes increased above
approximately $19\deg $, there is a sharp $(1-2^{\circ})$ transition
to a regime of dimples with a structural aspect similar to an
\textit{orange skin} (see Fig.~\ref{fig:samples}c). For steeper
slopes, we observe the progressive onset of a \textit{chevron pattern}
characterized by a well defined angle (see Fig.~\ref{fig:samples}d).
The cross-over region is indicated approximatively by a dashed line on
the phase diagram. The chevron pattern forms quickly (typically five
seconds) behind the receding liquid contact line. The rhomboid
elements characterizing this structure have a slightly rounded
downhill tip and a height profile like fish scales or roof tiles, i.e.
the sediment is thickest at the downhill tip, with a shallow decrease
uphill, and a sharp lower edge.

Finally, this regime is limited by the maximal stability angle
$\theta_{m}=35{{}^\circ}$ above which the sediment layer would
spontaneously avalanche as a whole.

\begin{figure}[htb!]%
  \includegraphics[width=0.9\columnwidth]{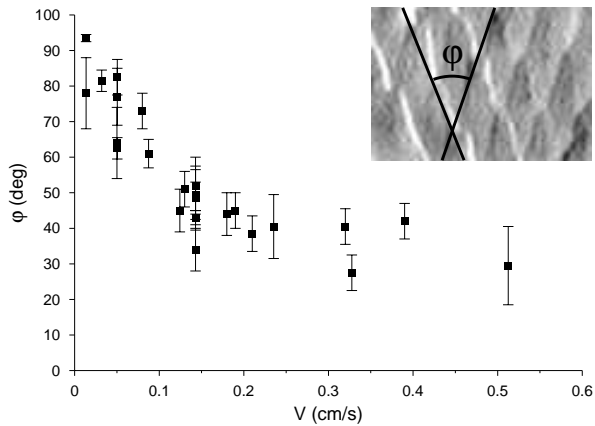}%
  \caption{Chevron alignment angle as a function of velocity. Error bars 
           indicate measurement variations}%
  \label{fig:chevrons_angle_V}%
\end{figure}

From the present experiment, we observe no systematic variation of the
chevron wavelength $\lambda $ with velocity or tilt angle.
Nevertheless, the spacing of the chevrons has a tendency to grow as
the contact line recedes with a mechanism akin to defect fusion; we
obtain a mean spacing $\lambda =5\,\mbox{mm}\pm 2\,\mbox{mm}$. On the
other hand, systematic experiments at constant angle $\theta =30\deg
$, show clearly a decrease of the chevron pattern opening angle $\phi
$ from $90\deg $ to $30\deg $ for increasing velocity (see
Fig.~\ref{fig:chevrons_angle_V}).

Various rhomboid patterns have already been described by geologists in
natural environments 
\cite{allen1984} like on sea-ward facing beach
slopes \cite{chevronbib},
in reservoirs or river beds after a full drainage of
the water. Although there seem to be several distinct types, only few
attempts at explaining a possible mechanism were made \cite{chevronbib}. Previous observations and the few
experiments available commonly attribute the formation of chevron
patterns to instabilities at the free surface of the flowing water
layer (such as hydraulic jumps) which couple to the bottom
profile~\cite{surfacewavesbib}. Although this could be true for
certain types of rhomboid ripples in shallow, fast flowing rivers, it
cannot be the case for the chevron regime described here. Indeed, the
largest estimation for the Froude number we can make at the chevron
onset is ${\rm Fr}=V/\sqrt{gd}=5\cdot 10^{-2}$, which certainly rules
out factors like hydraulic jumps of the water layer.
\begin{figure}[htb!]%
  \includegraphics[width=0.7\columnwidth]{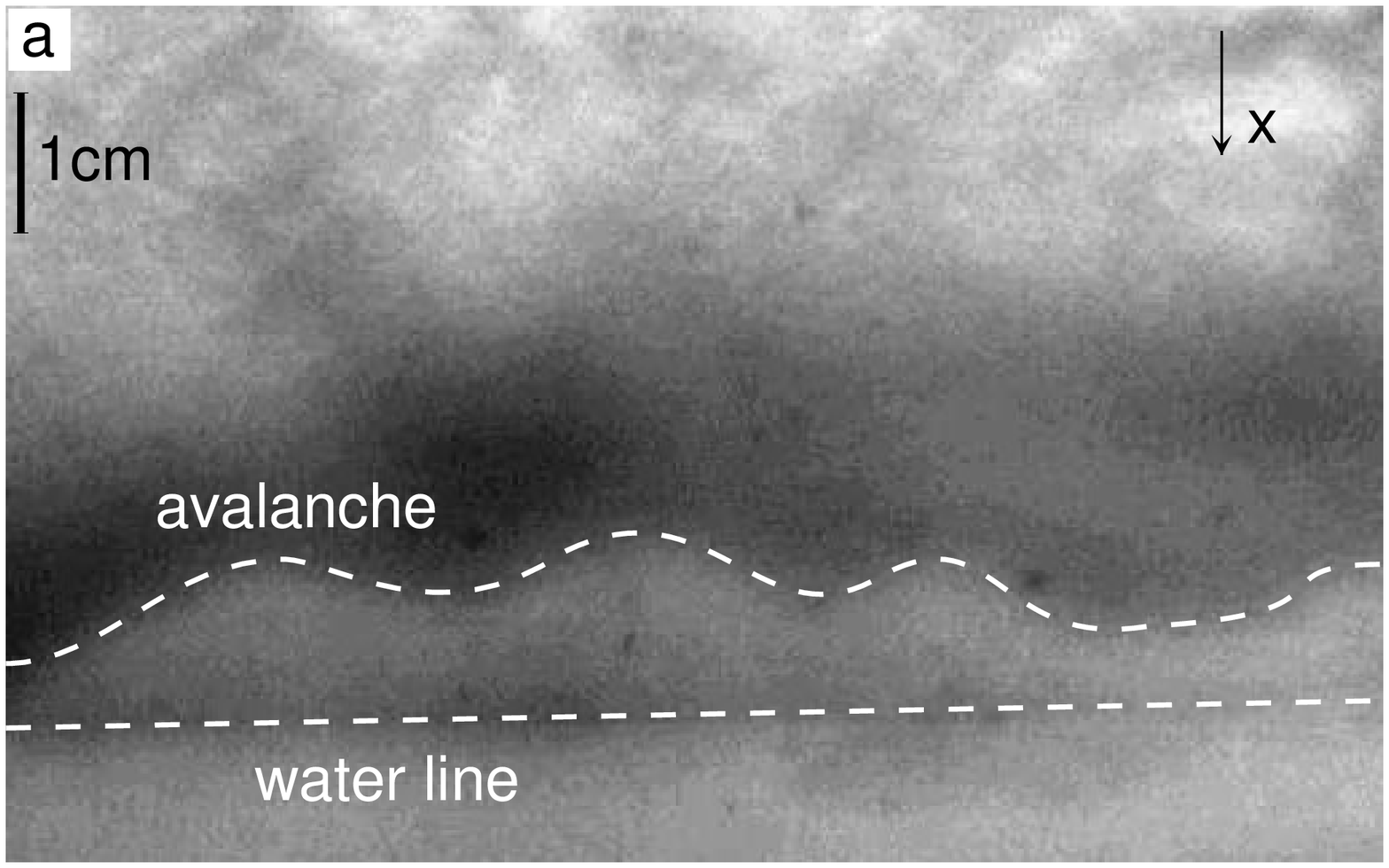}\\[2mm]%
  \includegraphics[width=0.7\columnwidth]{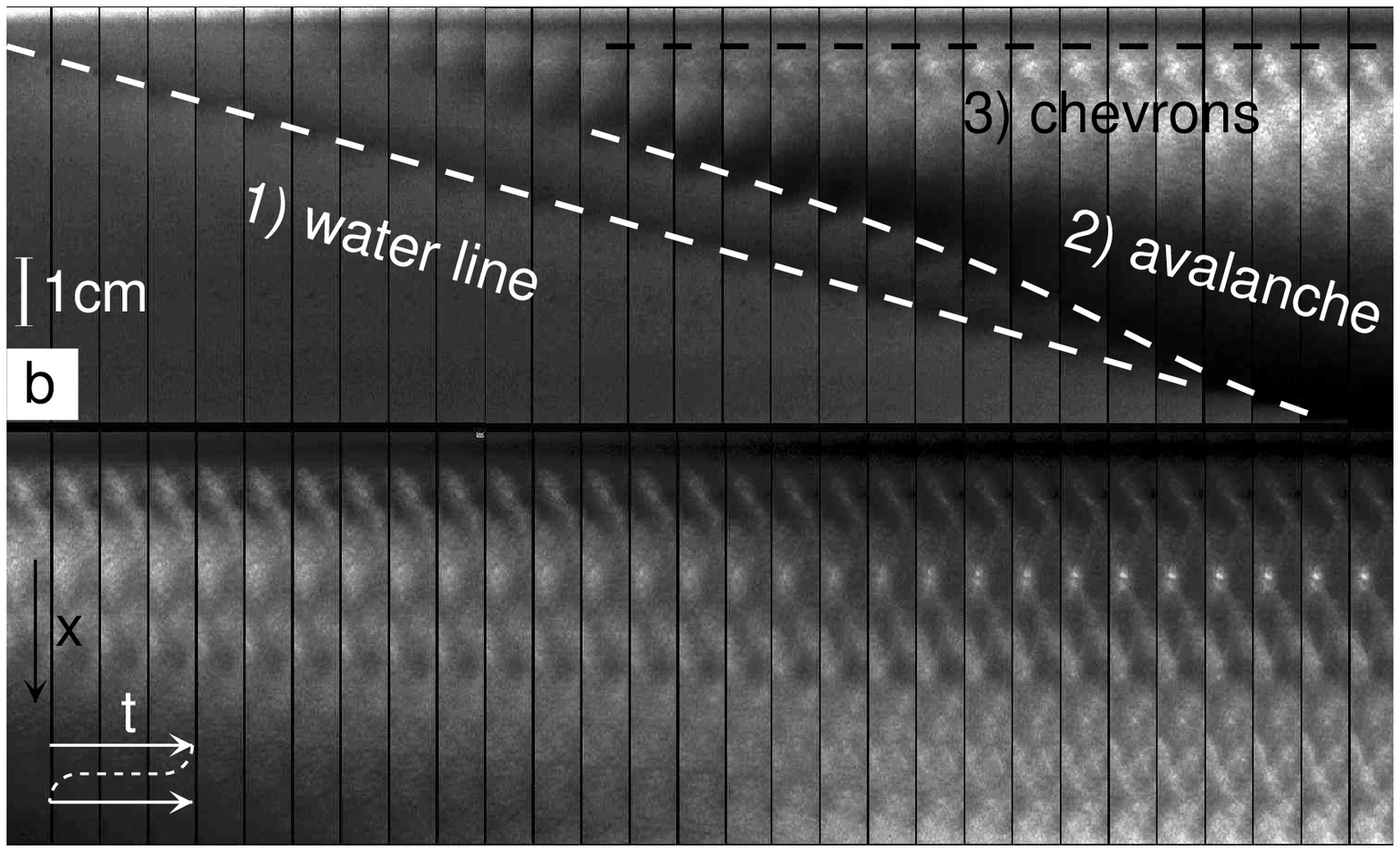}%
  \caption{Dynamics of the chevron formation. {\bf a} snapshot of the mud 
           avalanche above the contact line. {\bf b} spatio-temporal diagram 
           (split across two lines) showing how chevrons appear in the wake 
           of an avalanche.}%
  \label{fig:chevron_formation}%
\end{figure}%

Now we seek to clarify the physical conditions associated with the
onset of pattern formation. An estimation of the Darcy flow velocity
$V_{D}$ inside the powder yields $V_{D}=gK/\nu \simeq
10^{-5}\,\mbox{m/s}$, which is much smaller than the retrieval
velocity $V$. This $V_{D}$ value is obtained with a permeability
$K=10^{-12}\,\mbox{m}^{2}$ obtained experimentally, and a kinematic
viscosity $\nu =10^{-6}\,$m$^{2}$/s. Moreover, the capillary length
corresponding to a $30\,\mu$m porous medium under gravity forces is
about $20$\,cm. It is thus legitimate to consider that the sediment
remains fully soaked with water during retrieval of the plate. To
estimate the shear exerted by the liquid film at the surface, we
calculate its thickness $h(x,t)$ in the plate reference frame. The
retrieval of the plate begins at $t=0$, and $x=0$ is the initial
position of the contact line between liquid surface and sediment.
Assuming that the flow is viscous, so that the average local flow
velocity is given by $\overline{V}(x,t)=g\sin\theta
\,h(x,t)^{2}/3\nu$, the mass conservation equation $\partial_{t}h +
\partial_{x}\left( h(x,t)\overline{V}(x,t)\right) =0$ yields, for
small free flow slopes, a self consistent solution: $h(x,t)=\sqrt{2\nu
  x/tg\sin \theta}$. Note that we neglected both capillary and
hydrostatic pressure terms because of the small thickness and
curvature of the flowing layer. This approximation ceases to be valid
close to the origin and in the vicinity of the junction with the flat
water level. Note also that the contact line cannot move within this
approximation, which corresponds to a situation of total wetting on
the sediment. The maximum height, just above the reservoir water
level, is thus evaluated to be $h=\sqrt{(\nu /g\sin \theta
  )\,V}=10\,\mu \mbox{m}\approx d$. The first conclusion is that the
Reynolds number, ${\rm Re}=hV/\nu =0.03$, is small enough to justify
the lubrication approximation. Second, the ratio of the shear exerted
by the fluid on a grain at the bottom and its apparent weight yields
a common criterion for the onset of erosion called Shield's number,
\[
S=\frac{\rho _{w}\tan \theta }{\Delta \rho }\frac{h(x,t)}{d}\simeq \left( 
\frac{V}{V_{0}}\right) ^{1/2}\mbox{ with }V_{0}=\left( \frac{\Delta \rho }{%
\rho _{w}}\right) ^{2}\frac{gd^{2}}{\nu } 
\]
$\rho_{w}$ is the density of water and $\Delta\rho = \rho - \rho_{w}$ the
density contrast between grains and liquid.
On Fig.~\ref{fig:samples} we plotted the $V(\theta )$ curve
corresponding $S=0.12$. The scaling implications of this formula should
be put to test more systematically but so far, it seems to reproduce
remarkably the shape of the limit were erosion patterns are evidenced.
Note that a Shields number of value $S=0.12$ is marginally large to
represent a situation where a grain would be spontaneously dislodged
under the action of viscous shear. On the other hand, when interpreted
in the frame work of a Coulomb criterion for the sediment layer
stability, the shearing strength due to viscous forces could be large
enough to trigger an avalanche of wet grains. This is indeed what we
observe at an angle above around $18{{}^\circ} \pm 1{{}^\circ}$. For
this purpose, we visualized the experiment in the chevron regime by
lighting the flow through the bottom of the set-up
(Fig.~\ref{fig:chevron_formation}a): on this figure, darker spots
corresponding to thicker sediment layers. On
Fig.~\ref{fig:chevron_formation}b we present a space-time view of the
dynamical process for erosion and chevron formation. Each stripe of
actual size $0.7\times 6.1$ cm$^{2}$ corresponds to the central part
of a picture like 4a. Each successive stripe corresponds to a time lag
of $0.04s$. The line labeled by 1 indicates the water line retreating
at constant velocity. The line by labeled 2 shows the front of a
sediment avalanche triggered at the very top of the soaked layer.
Clearly this avalanche accelerates as grains accumulate at the front,
and it eventually catches up with the waterline. As the avalanche
proceeds, a spatially modulated pattern of settling sediment occurs in
the wake of the surge (line 3 indicates first chevrons) from the top
downwards. During this settlement, an horizontal structure is created
with a well defined spatial selection already visible in the vicinity
of the avalanche onset. This original structure will give rise to the
chevrons pattern since the subsequent sedimented structure will occur
further down with an horizontal spatial shift of roughly one half of
the wave length. Then a similar scenario is taking place for the next
structures down and so on. After final settlement of a chevron line,
sediment will still flow but only between adjacent chevrons. Now,
instead of an avalanche flow, we rather have a channel flow which
displays an erosive activity sharpening the pattern until it reaches
its final form. This final erosion process appears as a backwards
wave. At larger velocities the channel flow is sometimes able to erode
the layer down to the bottom of the plate, producing a river-like
network as in Fig.~\ref{fig:samples}e.

There are many open questions left on why such a structure is likely
to occur in the wake of the avalanche, and why it has such a
well defined wave-length. Here we mention two plausibly relevant paths
of thought. Recent experiments on sheared dense suspensions in a
rotating drum show that a particle dense flow is instable under shear
\cite{tirumkudulu1999}. Particle segregation occurs in the transverse
direction, producing regions denser in solid particles that may
enhance in our situation a localized sedimentary process.  Note also
that the dynamical aspects related to this mud avalanche bears many
similarities with studies on instabilities and pattern formation in
falling viscous sheets \cite{liu1995}. To our understanding, this
chevron regime reveals an interesting and open problem of mud flow and
sedimentation wave propagation. Along this line, it is interesting to
note that just at the onset of instability, in a tiny velocity region,
a regime of localized pulses crossing each other at $90\deg $ is
observed (see Fig.~\ref{fig:samples}f), which is actually strongly
reminiscent of a non-linear waves phenomenology.

In conclusion, we presented new results on a laboratory scale experiment
producing a rich variety of patterns due to erosion deposition processes.
The notion of ``phase'' we use in the scope of this manuscript is
qualitative (based on visual distinction), and should be defined more 
precisely in terms of an order parameter. We show that
contrarily to previous propositions these structures do not occur as a
consequence of water surface waves, but we postulate that they are
triggered by water viscous strain in low Reynolds and Froude numbers
regime. The corresponding Shield's criterion implies certain scalings of
the onset with grain size, viscosity etc., which are currently
investigated. We especially focussed here on an intriguing rhomboid-like
(chevron) structure organized with a typical distance and a coherent angle
between chevrons. This avalanche regime is clearly separated from another
erosion mechanism producing disordered networks of branched or
cross-hatched channels. In the chevron regime we evidence a new mechanism
based on the triggering of a dense sediment avalanche. The observed
patterns stems from instabilities initiated in the wake of the avalanche.
Erosion/deposition mechanisms further develop in order to carve the
pattern to its final chevron form. 


\noindent {\bf {\small Acknowledgments.}} We thank P.\ Gondret,
F.\ M\'{e}tivier and J.\ Socolar for interesting discussions, J.\ 
Treiner for photographs of natural erosion patterns. The help of G.\ 
Gu{\'{e}}na on the experiment is also acknowledged.

\bibliography{Abrv_Journaux,erosion,mouillage,local,daerr,livres}

\end{document}